\documentclass[aps,prb,twocolumn,superscriptaddress,showpacs]{revtex4-2}
\usepackage{graphicx}
\usepackage{amsmath}
\usepackage{amssymb}
\usepackage{hyperref}
\usepackage{dcolumn}
\usepackage{bm}
\usepackage[utf8x]{inputenc}
\usepackage{xcolor}
\begin{document}
\title{\texorpdfstring{Homogeneous to multi-scale inhomogeneous complex magnetism in Ba$_{3}$LnRu$_{2}$O$_{9}$ (Ln = Gd, Ho), \textcolor{black}{Revealed by $\mu$SR and Machine-Learned Muon-Site Calculations}}{Homogeneous to multi-scale inhomogeneous complex magnetism in Ba_3LnRu_2O_9 (Ln = Ho, Gd), Revealed by μSR and Machine-Learned Muon-Site Calculations}}

\author{S. Ghosh}
\email{sayan@rgipt.ac.in}
\affiliation{Rajiv Gandhi Institute of Petroleum Technology, Jais, Amethi, 229304, India}
\author{K. Sharma}
\affiliation{Rajiv Gandhi Institute of Petroleum Technology, Jais, Amethi, 229304, India}
\author{E. Kushwaha}
\affiliation{Rajiv Gandhi Institute of Petroleum Technology, Jais, Amethi, 229304, India}
\author{M. Kumar}
\affiliation{Rajiv Gandhi Institute of Petroleum Technology, Jais, Amethi, 229304, India}
\author{G. Roy}
\affiliation{Rajiv Gandhi Institute of Petroleum Technology, Jais, Amethi, 229304, India}
\author{A. Pandey}
\affiliation{Rajiv Gandhi Institute of Petroleum Technology, Jais, Amethi, 229304, India}
\author{F. L. Pratt}
\affiliation{ISIS Neutron and Muon Source, STFC, Rutherford Appleton Laboratory, Chilton, Oxon OX11 0QX, United Kingdom}
\author{S. Cottrell}
\affiliation{ISIS Neutron and Muon Source, STFC, Rutherford Appleton Laboratory, Chilton, Oxon OX11 0QX, United Kingdom}
\author{D. T. Adroja}
\affiliation{ISIS Neutron and Muon Source, STFC, Rutherford Appleton Laboratory, Chilton, Oxon OX11 0QX, United Kingdom}
\affiliation{Highly Correlated Matter Research Group, Physics Department, University of Johannesburg, Auckland Park 2006, South Africa}
\author{T. Basu}
\email{tathamay.basu@rgipt.ac.in}
\affiliation{Rajiv Gandhi Institute of Petroleum Technology, Jais, Amethi, 229304, India}

\begin{abstract}
We report a comprehensive muon spin rotation and relaxation ($\mu$SR) investigation of the hexagonal 6H-perovskites Ba$_{3}$LnRu$_{2}$O$_{9}$ \textcolor{black}{(Ln = Gd, Ho)} to understand the role of rare-earth anisotropy in governing the magnetic ground state of 4$d$-4$f$ coupled systems. Ba$_{3}$HoRu$_{2}$O$_{9}$ shows two successive magnetic orderings, $T_{N1} = 46$ K and $T_{N2} = 10$ K and a complex magnetic ground state; on the other hand, Ba$_{3}$GdRu$_{2}$O$_{9}$ consists of Gd$^{3+}$ ($L = 0$) S-state ions and shows primarily single magnetic ordering at $T_N = 14.5$ K. Zero-field (ZF) and longitudinal-field (LF) $\mu$SR measurements reveal fundamentally different magnetic behavior in the two compounds despite their structural similarity. Ba$_{3}$HoRu$_{2}$O$_{9}$ exhibits a multi-scale inhomogeneous magnetic state characterized by two-component exponential relaxation, broad local-field distributions, incomplete longitudinal-field decoupling, and persistent spin dynamics over a wide temperature range. The $\mu$SR results identify two characteristic magnetic energy scales near 46 and 10 K, associated with correlated Ru-Ho interactions and quasi-static magnetic ordering, respectively. Redfield analysis of the LF dependence of the relaxation rate further demonstrates the coexistence of strong local magnetic fields and distributed spin fluctuation timescales, confirming the presence of magnetic inhomogeneity. \textcolor{black} {Furthermore, machine-learning-assisted muon site and dipolar field calculations confirm that this broad field distribution naturally emerges from muons sensing distinct, coexisting magnetic domains across different structural sites.} In contrast, Ba$_{3}$GdRu$_{2}$O$_{9}$ shows a comparatively homogeneous antiferromagnetic ground state below 14.5 K, with Gaussian relaxation consistent with a narrow static field distribution. \textcolor{black} {Our results establish that varying the rare-earth ion in the Ba$_{3}$LnRu$_{2}$O$_{9}$ series acts as a powerful tuning factor using competing 4$d$-4$f$ magnetic forces to tune the ground state from homogeneous to a complex, multi-scale inhomogeneous state.}
\end{abstract}

\maketitle

\section{Introduction}

Transition-metal oxide-based strongly correlated electronic systems continue to attract significant interest because they host a variety of unusual electronic and magnetic states arising from the coupling between spin, charge, orbital, and lattice degrees of freedom \cite{Ref1, Ref2, Ref3, Ref4}. In particular, 4d and 5d systems often exhibit unconventional magnetic behavior due to the combined effects of electron correlation and strong spin-orbit coupling \cite{Ref5, Ref6, Ref7}. When rare-earth ions with localized 4f moments are introduced into these systems, the magnetic behavior becomes even more complex because of additional 4d-4f interactions and competing exchange pathways \cite{Ref8, Ref9}. Although 3d-4f compounds have been widely investigated, comparatively fewer studies have focused on 4d-4f correlated materials \cite{Ref10, Ref11, Ref12, Ref13, Ref14}. Among such systems, the hexagonal 6H-perovskite series Ba$_{3}$LnRu$_{2}$O$_{9}$ (Ln = Y, La-Lu) has emerged as an interesting platform for studying correlated magnetism \cite{Ref15, Ref16, Ref17}. These compounds display a wide variety of magnetic ground states depending on the nature of the rare-earth ion, reflecting the role of Ru-Ln hybridization and 4d-4f coupling \cite{Ref18, Ref19, Ref20, Ref21}. Their crystal structure contains face-sharing RuO$_{6}$ octahedra forming Ru$_{2}$O$_{9}$ dimers, which are linked through corner-sharing LnO$_{6}$ octahedra (shown in Fig. 1). This arrangement creates multiple competing magnetic interactions both within the Ru dimers and between the Ru and rare-earth moments \cite{Ref16, Ref18, Ref20}. In addition, the mixed-valence state of Ru $\sim$  +4.5 supports dimer-based magnetism and strengthens electronic correlations \cite{Ref16, Ref18}. Owing to these factors, the Ba$_{3}$LnRu$_{2}$O$_{9}$ family exhibits several intriguing phenomena such as magnetic frustration, molecular-like magnetism, magnetodielectric coupling, spin-driven ferroelectricity, and notable magnetocaloric behavior \cite{Ref15, Ref16, Ref20, Ref22, Ref23}.
\begin{figure}[ht!]
\centering
\includegraphics[width= \columnwidth,angle=0]{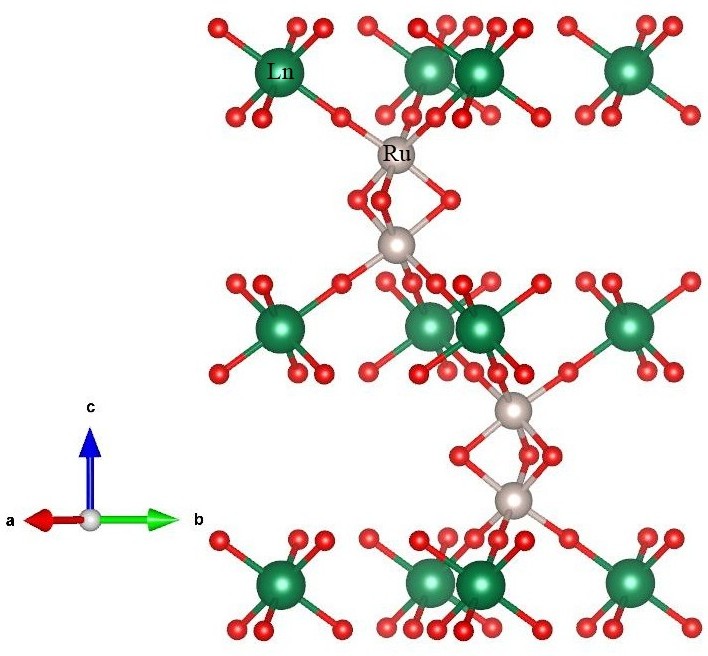}
\caption{\label{Fig-1} Crystal structure of the hexagonal 6H-perovskite Ba$_{3}$LnRu$_{2}$O$_{9}$ (Ln = Ho, Gd), showing face-sharing Ru$_{2}$O$_{9}$ dimers connected through corner-sharing LnO$_{6}$ octahedra. The geometry supports competing Ru-Ru and Ru-Ln magnetic interactions.}
\end{figure}

What makes this series particularly appealing is how sensitively its magnetic behavior depends on the choice of lanthanide ion. For non-magnetic lanthanide ions, the systems show a magnetic dimer behavior at high temperature \cite{Ref17}. However, the magnetic ground state of Ru varies sharply for different Ln-ions, due to change in LnO$_{6}$-octahedral distortion and results in different degrees of hybridization. Ba$_{3}$YRu$_{2}$O$_{9}$ exhibits a molecular magnetism with S = 1/2 dimer state as a result of strong metal-metal Ru-Ru bonding \cite{Ref24, Ref25}, whereas the isostructural Ba$_{3}$LaRu$_{2}$O$_{9}$ compound (La has a larger radius than Y) shows an orbital-selective state with S = 3/2 \cite{Ref25, Ref26}. For lighter rare-earth elements such as Nd, multiple magnetic transitions have been observed, involving both Ru and rare-earth sublattices \cite{Ref25, Ref27}. Interestingly, Nd shows ferromagnetic ordering, a contrasting behavior compared to any other members in this series \cite{Ref25, Ref27}. For heavier rare-earth ions, the picture initially appeared simpler, with a single antiferromagnetic transition around 10-15 K \cite{Ref16, Ref18, Ref20, Ref21, Ref22}. However, more recent studies have challenged this view. In particular, neutron diffraction results on Ba$_{3}$HoRu$_{2}$O$_{9}$ have revealed a much richer magnetic phase diagram, including an additional higher-temperature transition below 50 K with a cooperative ordering of Ho and Ru-moments, and the coexistence of competing magnetic phases at low temperatures \cite{Ref16, Ref18, Ref20}. Accompanied by spin-driven ferroelectricity below T$_{N2}$ = 10 K, these findings strongly suggest that the system is governed by competing interactions and multiple energy scales and raise the possibility that the magnetic state may be inhomogeneous in nature rather than uniformly ordered.

Bulk probes and neutron diffraction techniques are extremely powerful for identifying long-range magnetic order, but they often average out local variations and are less sensitive to slow spin dynamics or understanding of phase coexistence. A clear microscopic understanding is warranted to understand how magnetism develops locally in these materials and how competing magnetic phases coexist. This is precisely where muon spin relaxation ($\mu$SR) becomes particularly valuable. As a local probe, $\mu$SR directly senses the magnetic environment at the muon stopping site and can simultaneously detect static internal fields and slow spin fluctuations over a wide range of timescales \cite{Ref28, Ref29}. In systems where competing interactions lead to distributed or fluctuating magnetic states, $\mu$SR is uniquely suited to uncover signatures of inhomogeneity and multi-scale dynamics that may not be visible in other measurements \cite{Ref29, Ref30}.

In this work, we use zero-field (ZF) and longitudinal-field (LF) $\mu$SR to investigate the magnetic properties of Ba$_{3}$HoRu$_{2}$O$_{9}$ and ZF $\mu$SR to investigate Ba$_{3}$GdRu$_{2}$O$_{9}$. These two compounds offer a particularly useful comparison. Ho$^{3+}$ carries a strong orbital contribution and significant spin-orbit coupling, leading to pronounced magnetic anisotropy. In contrast, Gd$^{3+}$ is nearly isotropic (L = 0), providing a simpler reference system with minimal rare earth anisotropy. By studying these two materials side by side, we can directly assess how anisotropy and 4d-4f coupling influence both the static magnetic state and the underlying spin dynamics.

Our $\mu$SR results reveal a strong contrast between the two compounds. Ba$_{3}$HoRu$_{2}$O$_{9}$ shows an inhomogeneous magnetic ground state with coexistence of quasi-static magnetism and persistent spin dynamics, whereas Ba$_{3}$GdRu$_{2}$O$_{9}$ exhibits a comparatively homogeneous antiferromagnetic state. These findings demonstrate that rare-earth anisotropy and 4d-4f coupling play a key role in tuning multi-scale magnetism in the Ba$_{3}$LnRu$_{2}$O$_{9}$ family. \textcolor{black}{Furthermore, we have used machine-learning-assisted muon site calculation to understand the behavior in detail, a method which is of high interest at present in the application of physics.}

\section{Experimental details}

A polycrystalline sample of Ba$_{3}$HoRu$_{2}$O$_{9}$ was synthesized using a solid-state reaction method by mixing high-quality ($>$99.9\%) chemicals of BaCO$_{3}$, Ho$_{2}$O$_{3}$, and RuO$_{2}$, and to make Ba$_{3}$GdRu$_{2}$O$_{9}$, we used BaCO$_{3}$, Gd$_{2}$O$_{3}$, and RuO$_{2}$ as described in Ref. \cite{Ref21}. Measurements were made at the ISIS Neutron and Muon Source, Rutherford Appleton Laboratory, UK, to probe the microscopic magnetic properties and spin dynamics of these compounds. ZF $\mu$SR measurements on Ba$_{3}$HoRu$_{2}$O$_{9}$ were carried out using the EMU spectrometer \cite{Ref31}, whereas the measurements on Ba$_{3}$GdRu$_{2}$O$_{9}$ were performed on the HIFI spectrometer \cite{ref32}. In a $\mu$SR experiment, spin-polarized positive muons are implanted into the sample, where they sense the local magnetic field at the muon stopping site. The muons decay with a mean lifetime of 2.2 $\mu$s into a positron and two neutrinos. Since the emitted positron preferentially follows the direction of the muon spin at the moment of decay, the time evolution of the muon-spin polarization can be monitored through positron detection \cite{Ref28, Ref33}.

The asymmetry signal was obtained from the positron counts recorded in the forward (N$_{F}$) and backward (N$_{B}$) detector groups with respect to the initial muon-spin polarization direction using \cite{Ref29, Ref30},

\begin{equation}
A(t) = A_{0}P(t) =  \frac{N_F - \alpha N_B}{N_F + \alpha N_B}
\end{equation}

where $A_{0}$ is the full asymmetry for the spectrometer. $\alpha$ is a calibration factor accounting for the relative detector efficiency and geometrical asymmetry between the forward and backward detector pairs. The estimated values of $\alpha$ are 0.984 for Ba$_{3}$HoRu$_{2}$O$_{9}$ and 1.256 for Ba$_{3}$GdRu$_{2}$O$_{9}$, determined from transverse-field measurements with a 20 G field in the diamagnetic mode. The measured asymmetry is directly related to the muon-spin polarization. The $\mu$SR data analysis was carried out using the WiMDA software package \cite{Ref34}.

\textcolor{black}{We have used machine learning-based MEOWN software to determine the muon stopping sites and the electric dipolar fields, which complement our experimental findings \cite{ref35}.}

\section{Results \& discussion}

The ZF $\mu$SR measurements reveal the essential differences between the two systems, Ba$_{3}$HoRu$_{2}$O$_{9}$ and Ba$_{3}$GdRu$_{2}$O$_{9}$. In Fig. 2a, the ZF asymmetry spectra of Ba$_{3}$HoRu$_{2}$O$_{9}$ are well fitted using a two-component exponential relaxation function,

\begin{equation}\label{eq-2}
A(t) = A_1 e^{-\lambda_1 t} + A_2 e^{-\lambda_2 t} + A_{bg}
\end{equation}

Where A$_{1}$ and A$_{2}$ are the initial asymmetries, $\lambda$$_{1}$ and $\lambda$$_{2}$ are the relaxation rates, and A$_{bg}$ is the temperature-independent background contribution arising mainly from muons stopping in the silver holder or outside the sample. In the fitting we got, A$_{bg}$ = 0.907 \%. The presence of two relaxation components indicates that the implanted muons experience a broad range of local magnetic environments. In $\mu$SR, such behavior generally appears when the internal magnetic fields are widely distributed, either because of spatial magnetic inhomogeneity or because different regions fluctuate at different rates or have multiple muon sites. The fact that these two relaxations remain present throughout the entire temperature range suggests that the magnetic state is heterogeneous, with different magnetic regions evolving on different timescales.

\begin{figure}[!ht]
\centering
\includegraphics[width=3.5in,angle=0]{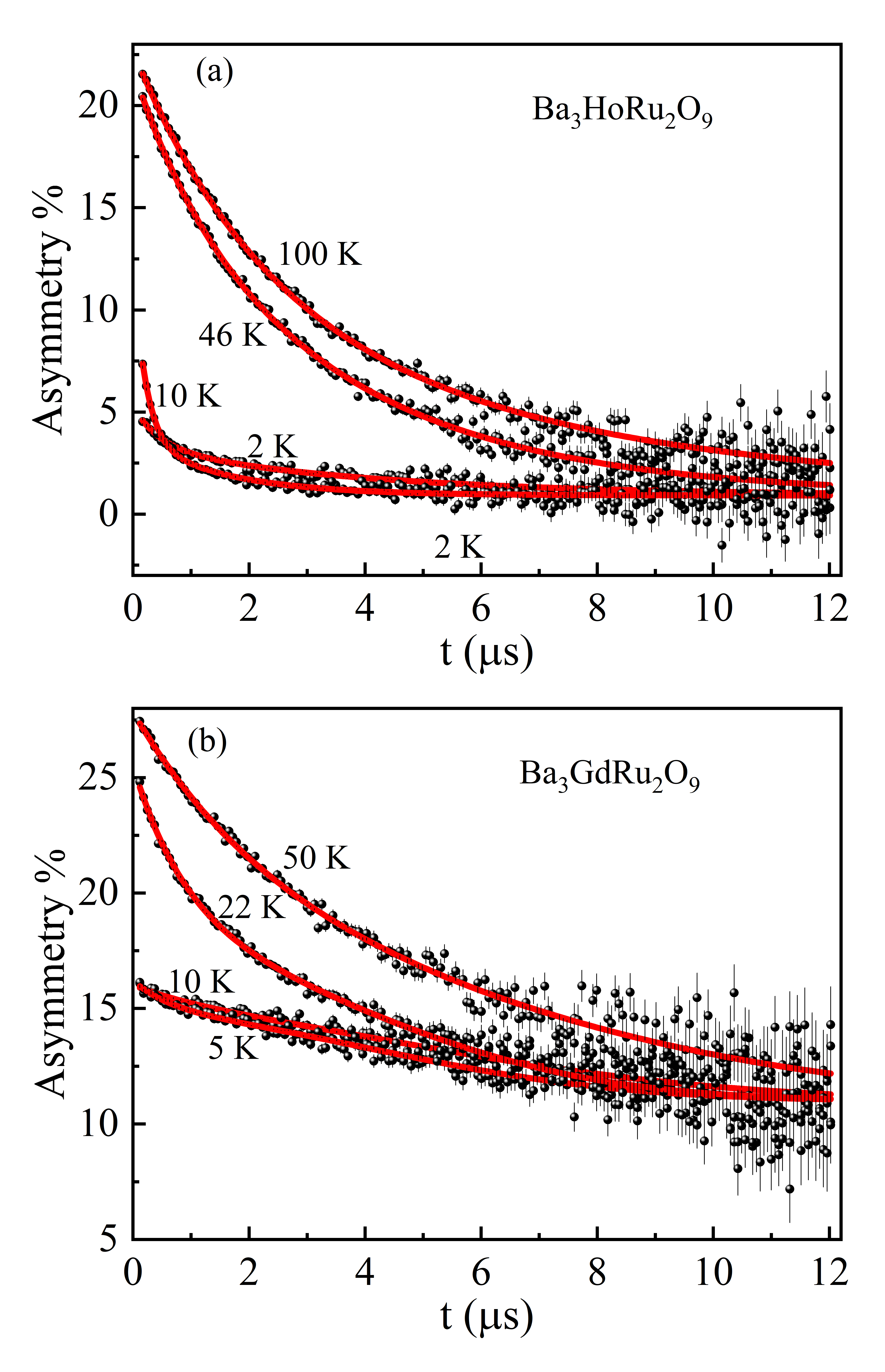}
\caption{\label{Fig-2} ZF $\mu$SR asymmetry spectra at selected temperatures for (a) Ba$_{3}$HoRu$_{2}$O$_{9}$ and (b) Ba$_{3}$GdRu$_{2}$O$_{9}$. Solid lines are fits using Eq. (2) and Eq. (3), respectively. The Ho compound shows two-component exponential relaxation, whereas the Gd compound exhibits Gaussian-like relaxation consistent with a more homogeneous magnetic state.}
\end{figure}

Fig. 3a shows that both asymmetry components exhibit nearly a 2/3 drop near T$_{N2}$ $\sim$ 10 K, indicating nearly long-range magnetic order. In Fig. 3b, both the fast relaxation rate $\lambda$$_{1}$ and the slow relaxation rate $\lambda$$_{2}$ increase strongly around 10 K. This enhancement reflects the development of quasi-static internal magnetic fields at the muon site and agrees well with the magnetic transition at 10 K. However, no oscillation in the asymmetry spectra is observed in the ordered state. In a magnetically ordered system with a well-defined internal field, coherent muon spin precession is normally expected. The absence of oscillations suggests that the internal fields are broadly distributed, pointing toward multiple local magnetic configurations. It is also possible that the oscillation frequency lies beyond the time resolution of the ISIS pulse width, as these compounds have large internal fields. The broad maxima observed in both $\lambda$$_{1}$ (T) and $\lambda$$_{2}$ (T) near 10 K further indicate critical slowing down of spin fluctuations close to the ordering temperature.

\begin{figure}[ht!]
\centering
\includegraphics[width=\columnwidth,angle=0]{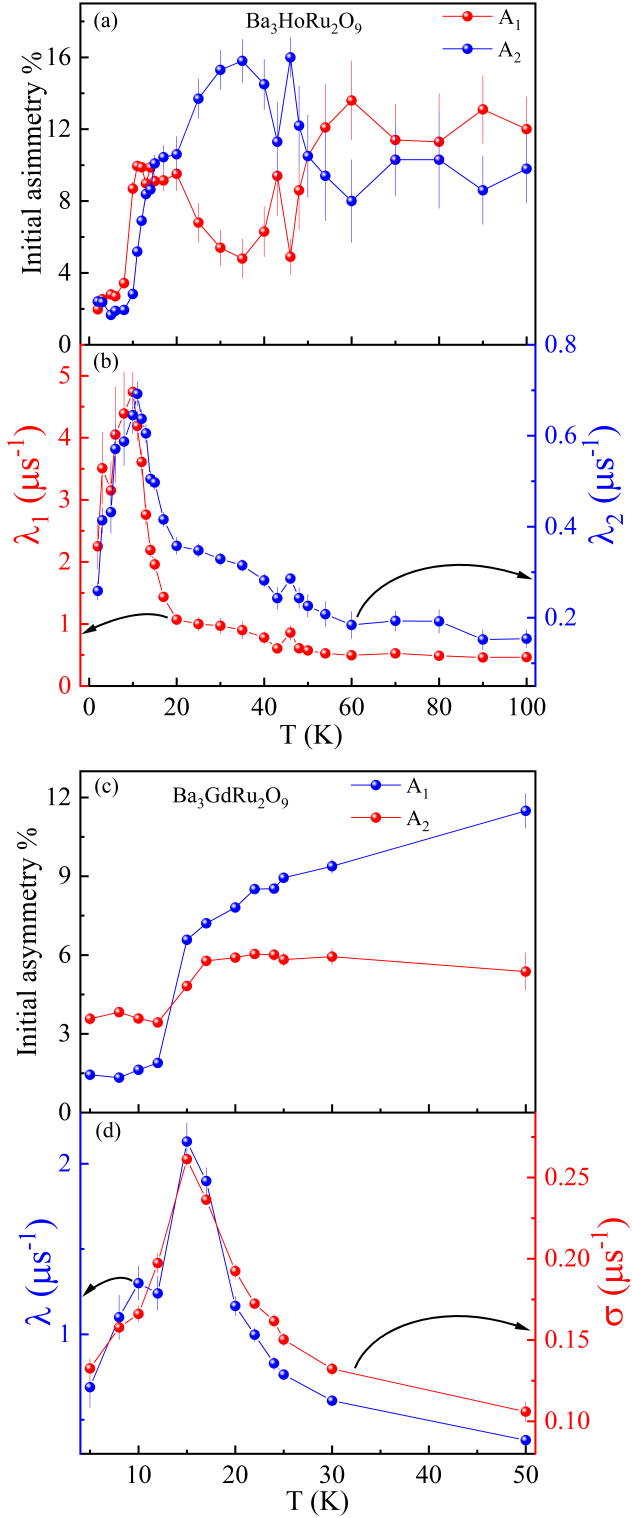}
\caption{\label{Fig-3} Temperature dependence of the ZF $\mu$SR fitting parameters. (a,b) Initial asymmetries and relaxation rates for Ba$_{3}$HoRu$_{2}$O$_{9}$, showing magnetic anomalies near 10 K and persistent correlations up to 46 K. (c,d) Corresponding parameters for Ba$_{3}$GdRu$_{2}$O$_{9}$, with a clear peak near 14.5 K associated with antiferromagnetic ordering.}
\end{figure}

Another important result is that finite relaxation remains visible well above the low-temperature transition and persists up to nearly T$_{N1}$ $\sim$ 46 K, close to the previously reported transition below 50 K \cite{Ref18}. This shows that magnetic correlations survive far above the main ordering temperature and that the system does not simply change from a paramagnetic state to a fully ordered one. Instead, the correlations develop gradually over a broad temperature range, indicating multiple magnetic energy scales. The higher temperature scale around 46 K is likely related to correlated interactions between Ru dimers and Ho moments and their cooperative order, whereas the lower transition near 10 K corresponds to the appearance of quasi-static magnetism. The persistence of two relaxation components across the full temperature range further confirms that the system maintains a wide distribution of local magnetic environments and fluctuation timescales even above the ordering region T$_{N2}$ and T$_{N1}$.

In contrast, the ZF $\mu$SR spectra of Ba$_{3}$GdRu$_{2}$O$_{9}$ show a very different behavior, indicating a much more uniform magnetic state. The spectra are well fitted using a combination of exponential and Gaussian relaxation functions,
\begin{equation}
A(t) = A_1 e^{-\lambda t} + A_2 e^{-\sigma^2 t^2 / 2} + A_{bg}
\end{equation}
where the Gaussian term represents a static distribution of local magnetic fields and the exponential term describes dynamic fluctuations (shown in Fig. 2b). Here, A$_{1}$ and A$_{2}$ are the initial asymmetry parameters, while $\lambda$ and $\sigma$ are the corresponding muon relaxation rates. The fitting gives A$_{bg}$ = 11 \%, \textcolor{black}{which is relatively large and arises from the muons implanted outside of the sample or from the silver sample holder. This background will not affect much, as we are getting all the features reasonably well}. The Gaussian relaxation is characteristic of randomly oriented static fields with a relatively narrow field distribution. At low temperatures, this Gaussian contribution becomes dominant, indicating the formation of a comparatively homogeneous static magnetic state, consistent with conventional antiferromagnetic ordering of the Gd$^{3+}$ moments.

\begin{figure}[!ht]
\centering
\includegraphics[width=3.5in,angle=0]{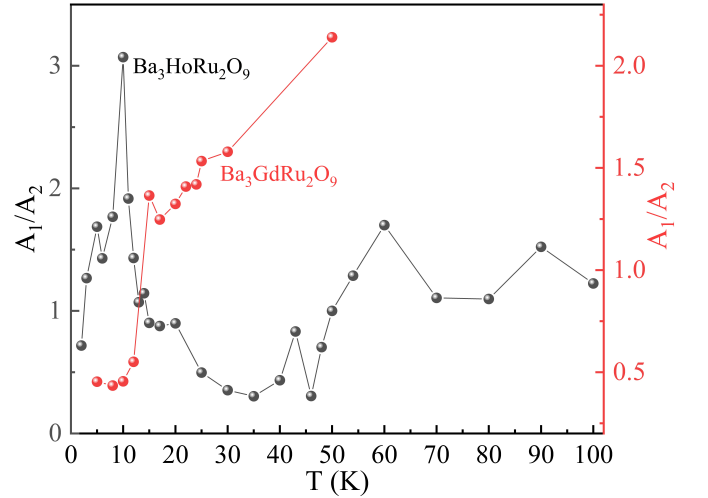}
\caption{\label{Fig-4} \textcolor{black}{Temperature dependence of the initial asymmetry ratio $A_1/A_2$ for Ba$_{3}$HoRu$_{2}$O$_{9}$ (black spheres, left axis) and Ba$_{3}$GdRu$_{2}$O$_{9}$ (red spheres, right axis).}}
\end{figure}

The temperature dependence of the initial asymmetry also supports this picture. In Fig. 3c, a clear reduction of the initial asymmetry is observed below 14.5 K \cite{Ref21}. The slowly relaxing component increases above this temperature, while the fast relaxing component dominates at low temperatures. This behavior reflects progressive depolarization of the muon spins due to the development of static internal magnetic fields below the ordering temperature. The smooth variation of the asymmetry suggests that the magnetic transition develops continuously rather than abruptly and remains consistent with a homogeneous magnetic phase.

Additionally, the temperature dependence of the muon relaxation rates is shown in Fig. 3d. The exponential relaxation rate displays a pronounced peak near 14.5 K, which is a typical sign of critical slowing down of spin fluctuations close to the magnetic transition. Below this temperature, the Gaussian fraction of the relaxing asymmetry increases below 14.5 K, indicating the development of static local magnetic fields in the ordered state. At higher temperatures, the exponential relaxation decreases gradually as the system enters a dynamic regime where spin fluctuations become faster than the $\mu$SR time window. Altogether, the coexistence of Gaussian and exponential relaxation gives a consistent picture of a crossover from a dynamic paramagnetic state at high temperature to a homogeneous static magnetic state at low temperature.
\begin{figure*}[htbp]
  \centering
  \includegraphics[width=\textwidth]{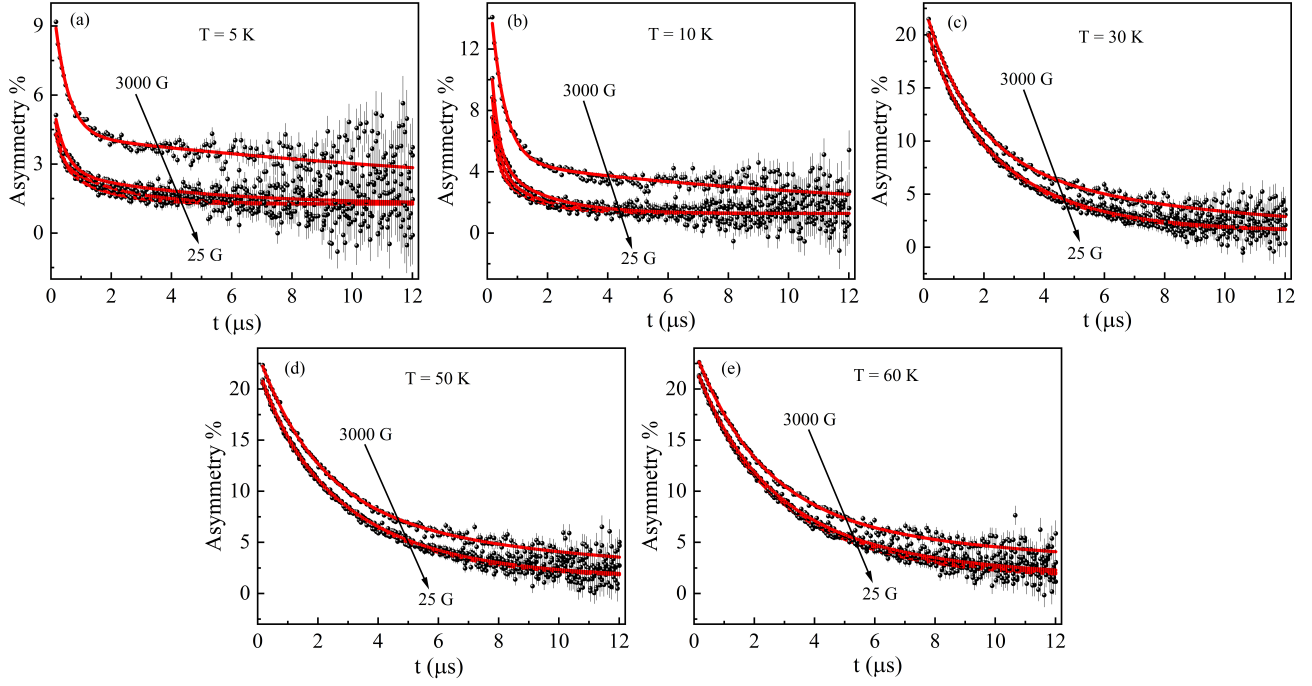}
  \caption{LF $\mu$SR asymmetry spectra of Ba$_{3}$HoRu$_{2}$O$_{9}$ measured under different longitudinal fields at (a) 5 K, (b) 10 K, (c) 30 K, (d) 50 K, and (e) 60 K. Solid lines are fits using \textcolor{black}{Eq. (2)}. Partial LF decoupling and remaining relaxation indicate coexistence of quasi-static and dynamic magnetism.}
  \label{Fig-4}
\end{figure*}
\ \\
\begin{figure*}[htbp]
  \centering
  \includegraphics[width=\textwidth]{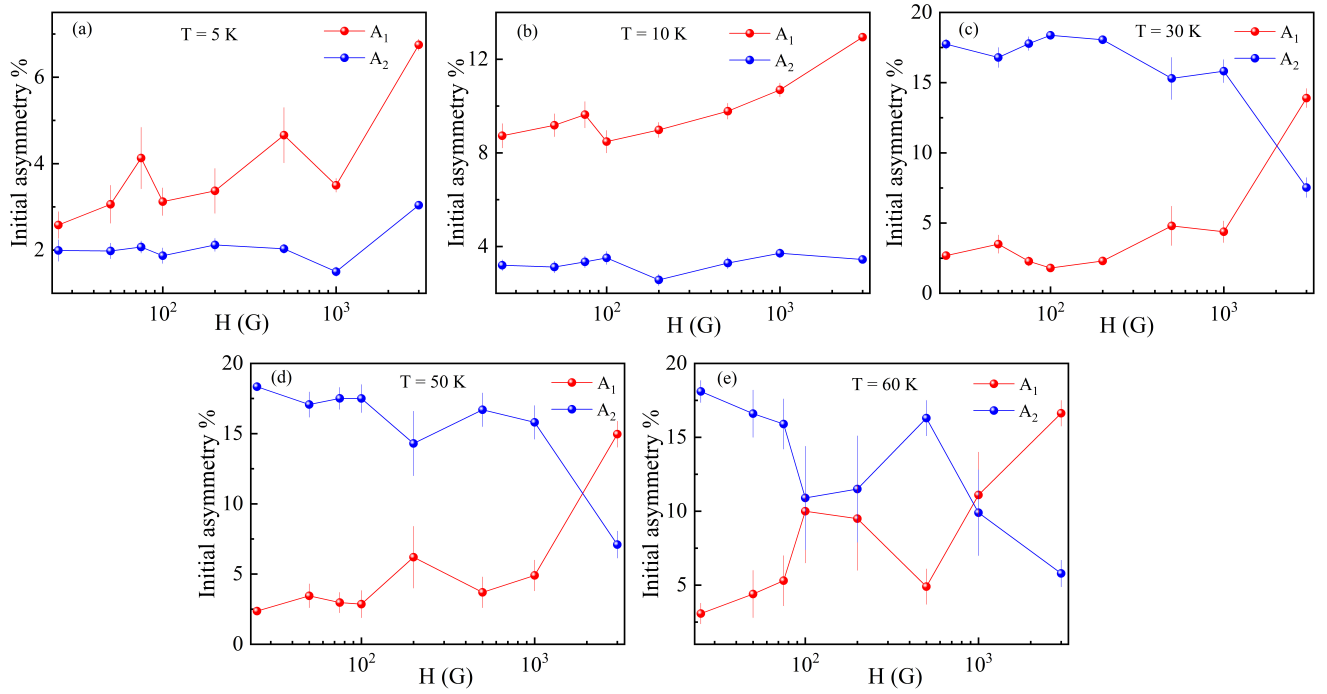}
  \caption{Field dependence of the asymmetry components A$_{1}$ and A$_{2}$ for Ba$_{3}$HoRu$_{2}$O$_{9}$ at different temperatures. The redistribution of spectral weight with field reflects mixed static and dynamic magnetic regions.}
  \label{Fig-5}
\end{figure*}
The difference between the two systems becomes especially clear when comparing their relaxation functions. Ba$_{3}$GdRu$_{2}$O$_{9}$ shows a Gaussian component, which is typical of a narrow and nearly isotropic static field distribution. On the other hand, Ba$_{3}$HoRu$_{2}$O$_{9}$ exhibits only exponential relaxation, indicating a much broader and more complicated distribution of local magnetic fields and fluctuation rates. This reflects the fundamentally different magnetic nature of the two compounds and highlights the strong role of anisotropy and competing magnetic interactions in the Ho-based system.

\textcolor{black}{Further, to test if the two muon signals are just a simple structural effect, we can look at the intensity ratio of the two components, $A_1/A_2$, shown in Fig. 4. As for Ba$_{3}$HoRu$_{2}$O$_{9}$, the weight of $A_1$ is larger than $A_2$ below LRO, If the magnetic state is purely static and strictly tied to the crystal geometry, this ratio should remain locked at 2:1. As the data clearly shows, Ba$_{3}$HoRu$_{2}$O$_{9}$ completely violates this rule; its $A_1/A_2$ ratio fluctuates wildly and deviates heavily from the 2:1 expectation below its LRO temperature, $T_{N2}$. This strong temperature dependence confirms that the signals in the Ho compound are driven by genuine magnetic phase coexistence, which was predicted in previous neutron studies \cite{Ref18}, and changing spin dynamics rather than a static geometric artifact. In sharp contrast, for Ba$_{3}$GdRu$_{2}$O$_{9}$, the weight of $A_2$ is larger than $A_1$ below LRO, the $A_1/A_2$ ratio should give 1:2 for static ground state. It drops smoothly and settles right around 0.5 below its magnetic ordering temperature of 14.5~K. This perfect match with the 1:2 geometric expectation confirms that the Gd compound behaves exactly as expected for a simple, homogeneous, and static magnetic ground state.}

\begin{figure*}[htbp]
  \centering
  \includegraphics[width=\textwidth]{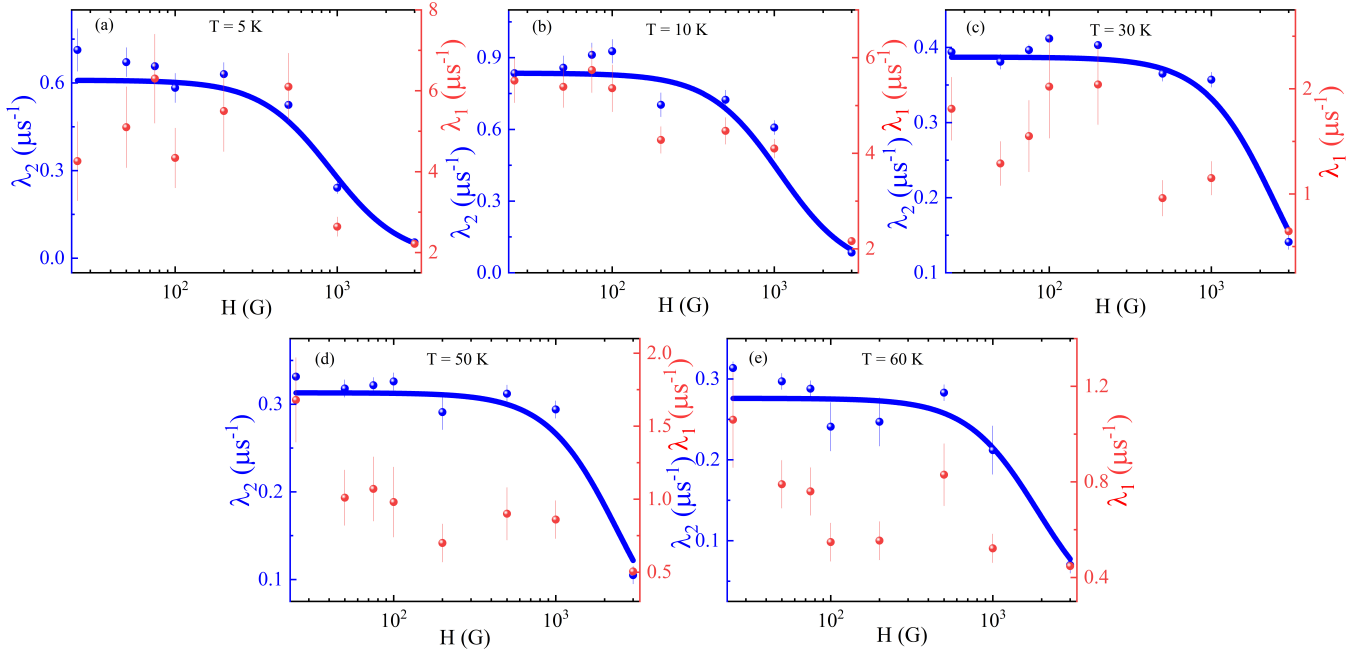}
  \caption{Field dependence of the relaxation rates $\lambda$$_{1}$ and $\lambda$$_{2}$ for Ba$_{3}$HoRu$_{2}$O$_{9}$. The solid blue curves are Redfield fits to $\lambda$$_{2}$(H). The incomplete suppression of relaxation under LF indicates distributed spin dynamics and magnetic inhomogeneity.}
  \label{Fig-6}
\end{figure*}

The LF $\mu$SR measurements on Ba$_{3}$HoRu$_{2}$O$_{9}$ provide further evidence for magnetic inhomogeneity and persistent spin dynamics. The LF spectra at different temperatures were analyzed using the same two-exponential model (Eq. 2) used for the ZF data with the value of A$_{bg}$ = 1.5 \%. From the fitting, $\lambda$$_{1}$ always remains larger than $\lambda$$_{2}$, showing that the A$_{1}$ component corresponds to the fast-relaxing channel, while A$_{2}$ represents the slower relaxing magnetic component. The field dependence of these parameters is presented in Figs.~\ref{Fig-5} and~\ref{Fig-6}.

At 5 K, well below the magnetic ordering temperature, the spectra show a clear redistribution between the two relaxation channels with increasing field. As seen in Fig. 5a, the amplitude of the fast-relaxing component A$_{1}$ increases gradually with LF, while the slow component A$_{2}$ decreases correspondingly. At the same time, the slow relaxation rate $\lambda$$_{2}$ is strongly suppressed with increasing field (shown in Fig. 6a), especially above about 1000 G. This behavior indicates progressive decoupling of muon spins from quasi-static internal magnetic fields and confirms that a large part of the magnetic field distribution becomes quasi-static at low temperature, consistent with long-range magnetic ordering. However, the decoupling remains incomplete even at the highest applied fields. Both $\lambda$$_{1}$ and $\lambda$$_{2}$ stay finite over the entire field range, and residual relaxation survives even under several kG of LF. In a fully homogeneous static magnetic state, nearly complete decoupling would normally be expected at sufficiently high field. The remaining relaxation, therefore, shows that dynamic magnetic fluctuations survive deep inside the ordered phase. This means that the low-temperature state of Ba$_{3}$HoRu$_{2}$O$_{9}$ is not uniformly static. Instead, static and fluctuating magnetic regions coexist, giving rise to an inhomogeneous magnetic ground state.

At 10 K (Figs. 5b and 6b), a similar but slightly weaker behavior is observed. The fast-relaxing amplitude A$_{1}$ continues to increase with field, whereas A$_{2}$ decreases gradually. At the same time, $\lambda$$_{2}$ decreases continuously under LF, confirming that static internal fields still don't dominate close to the lower magnetic transition. Nevertheless, neither relaxation rate vanishes completely, again showing coexistence of static and dynamic magnetic contributions. This suggests that magnetic moment in different domains develops gradually and unevenly, with different magnetic regions slowing down on different timescales.

At 30 K (Figs. 5c and 6c), the field dependence becomes much weaker compared to the low-temperature ordered state. Although A$_{1}$ and A$_{2}$ still change with the field, the suppression of $\lambda$$_{2}$ is much smaller and finite relaxation persists over the whole field range. This indicates that dynamic spin fluctuations become more dominant in the $\mu$SR response. At the same time, the partial LF decoupling and remaining relaxation show that short-range magnetic correlations or spin reorientations are still present, meaning that correlated and spatially inhomogeneous spin dynamics continue to exist at 30 K.

At 50 and 60 K (Figs. 5d,e and 6d,e), the LF dependence becomes even weaker, characteristic of a mainly dynamic regime consistent with the paramagnetic state. The fast-relaxing component A$_{1}$ becomes dominant at high fields, while A$_{2}$ decreases steadily. The relaxation rates are much lower than those at low temperatures and show only gradual suppression with the field instead of sharp decoupling. This indicates that the internal fields fluctuate rapidly on the $\mu$SR timescale. Both relaxation components remain finite at high fields, showing that correlated spin fluctuations survive well above the long-range ordering temperature near 46 K. This suggests the presence of fluctuating magnetic clusters or short-range correlated regions extending to high temperatures.

Collectively, the LF $\mu$SR measurements demonstrate that Ba$_{3}$HoRu$_{2}$O$_{9}$ does not evolve into a simple homogeneous static magnetic phase. Instead, the system shows coexistence of quasi-static and dynamic magnetic components over a broad temperature range. Strong LF decoupling at low temperatures confirms the formation of static internal fields due to long-range order, while the incomplete recovery of asymmetry and the remaining relaxation reveal the survival of slow spin fluctuations inside the ordered state. These results strongly support an inhomogeneous magnetic ground state in Ba$_{3}$HoRu$_{2}$O$_{9}$.
\begin{figure}[ht!]
\centering
\includegraphics[width=\columnwidth,angle=0]{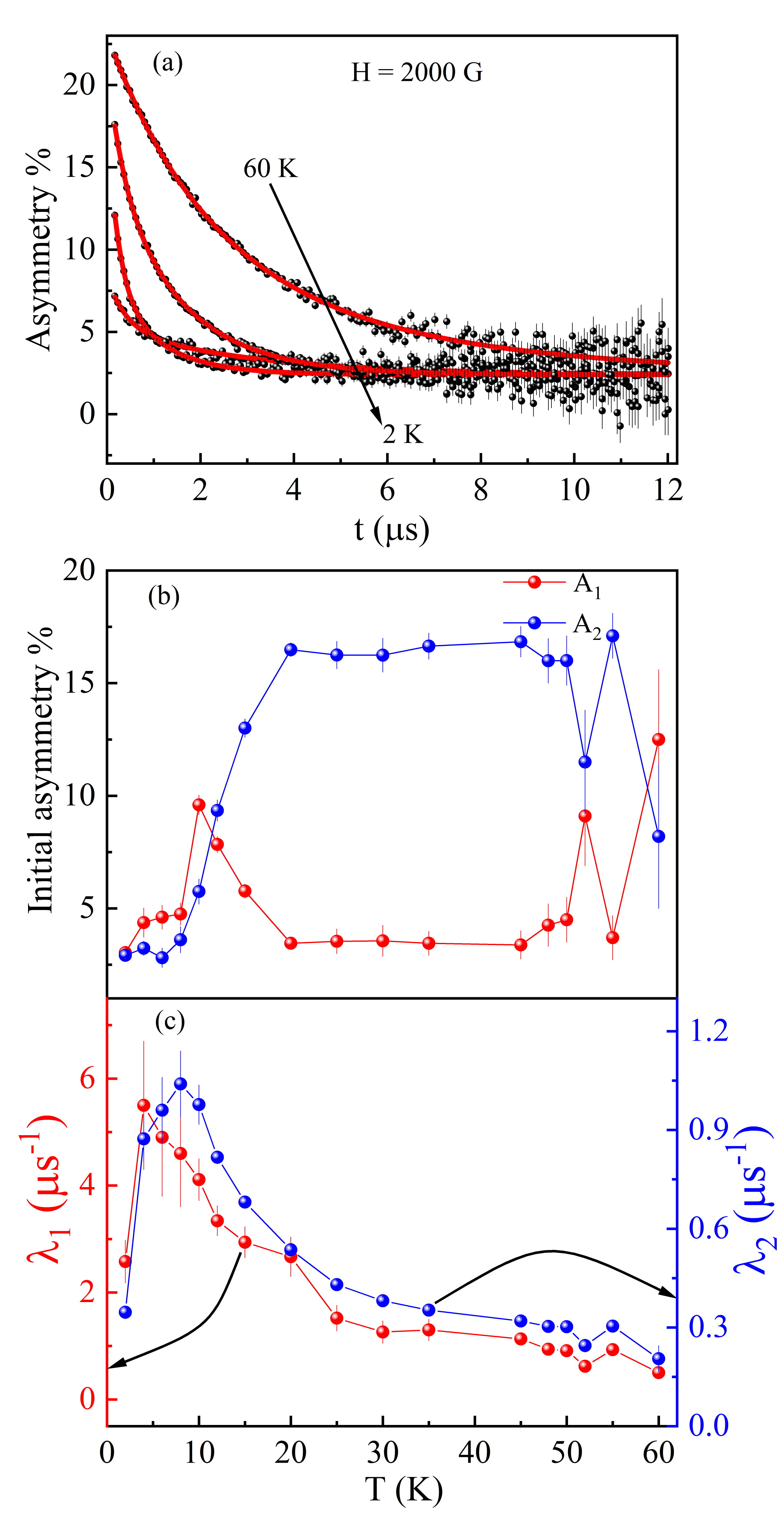}
\caption{\label{Fig-7} (a) LF $\mu$SR spectra of Ba$_{3}$HoRu$_{2}$O$_{9}$ measured under a fixed field of 2000 G at selected temperatures. Solid lines are fits using \textcolor{black}{Eq. (2)}. Temperature dependence of (b) asymmetry components and (c) relaxation rates. Persistent relaxation under 2000 G confirms the coexistence of quasi-static magnetism and slow spin fluctuations.}
\end{figure}
An important aspect of LF $\mu$SR is that the relaxation rate reflects the spectral density of spin fluctuations at the frequency $\omega\; = \gamma_{\mu}\text{H}$. Therefore, the field dependence of the relaxation directly probes the nature of local magnetic fluctuations and how they evolve with temperature.

To understand the spin dynamics quantitatively, the field dependence of the slow relaxation component $\lambda_{2}$ was analyzed using the Redfield theoretical model \cite{Ref36},

\begin{equation}
\lambda(H) = \frac{2 \gamma_\mu^2 H_{loc}^2 \tau_c}{1 + (\gamma_\mu H \tau_c)^2}
\end{equation}

where H$_{loc}$ is the local magnetic field at the muon site, and $\tau_{c}$ is the spin correlation time. The fits are shown in Fig. 6(a-e). The analysis was carried out only for $\lambda$$_{2}$ because this component exhibits clear, systematic LF decoupling across the entire field range. In contrast, $\lambda$$_{1}$ shows weak and irregular field dependence and therefore cannot be fitted reliably within the simple Redfield approach.

The Redfield analysis clearly reveals (shown in Table 1) that the local field H$_{loc}$ remains quite large at low temperatures, with values of about 1488 G at 5 K, 1412 G at 10 K, and 1102 G at 30 K. Such large internal fields further confirm strong quasi-static magnetic correlations in the ordered state. More importantly, H$_{loc}$ does not disappear abruptly above the magnetic transition. Even at 50 and 60 K, the extracted values remain sizeable, around 842 G and 781 G, respectively. This clearly demonstrates that short-range magnetic correlations survive far above the long-range ordering temperature. At the same time, the correlation time $\tau_{c}$ decreases steadily with increasing temperature, decreasing from about 0.0215 $\mu$s at 5 K to about 0.0053 $\mu$s at 60 K. This reflects the gradual acceleration of spin fluctuations as thermal energy weakens the quasi-static magnetic state. Consistently, the fluctuation frequency $\nu$ increases from about 46.5 MHz at 5 K to about 188.6 MHz at 60 K, indicating a crossover from slow correlated spin dynamics at low temperature to a much faster fluctuating regime at higher temperature.

\begin{table}[!h]
\centering
\begin{tabular}{ccccc}
\hline\hline
T (K) & H$_{loc}$ (G) for $\lambda$$_{2}$ & $\tau_{c}$ ($\mu$s) & $\nu$ (MHz) & State \\
\hline
5 & 1488.2 & 0.0215 & 46.5 & Ordered \\
10 & 1412.5 & 0.0125 & 80.0 & Ordered (T$_{N2}$) \\
30 & 1102.4 & 0.0061 & 163.9 & Spin reorientation \\
50 & 842.1 & 0.0055 & 181.8 & Paramagnetic \\
60 & 781.4 & 0.0053 & 188.6 & Paramagnetic \\
\hline\hline
\end{tabular}
\caption{Parameters obtained from the Redfield analysis of the LF $\mu$SR data for Ba$_{3}$HoRu$_{2}$O$_{9}$,}
\end{table}

An important outcome of this analysis is that the system simultaneously maintains strong local magnetic fields and finite spin dynamics over a wide temperature range. This behavior is inconsistent with a homogeneous magnetic state. Instead, it suggests that different magnetic regions or spin clusters fluctuate on different timescales, leading to distributed magnetic dynamics throughout the system. Together with the incomplete LF decoupling and the two-component relaxation observed in ZF $\mu$SR, the Redfield analysis strongly supports an inhomogeneous magnetic ground state in Ba$_{3}$HoRu$_{2}$O$_{9}$, where static and dynamic magnetic correlations coexist over a broad temperature range.

To get further insight into the magnetic ground state and spin dynamics of Ba$_{3}$HoRu$_{2}$O$_{9}$, we performed LF $\mu$SR measurements under a fixed longitudinal field of 2000 G, shown in Fig. 7a-c. Since this field is comparable to the characteristic internal field obtained from the Redfield analysis, these measurements effectively probe the competition between quasi-static and dynamic magnetic components over a wide temperature range.

We have fitted this asymmetry data with Eq.~\ref{eq-2}, where the A$_{bg}$ is calculated to be $\sim$ 2.4 \%. As shown in Fig. 7a, the $\mu$SR spectra evolve systematically with temperature. However, even at 2 K, the asymmetry does not recover completely under 2000 G. The remaining strong relaxation at short times indicates that part of the internal field distribution either stays quasi-static or fluctuates slowly within the $\mu$SR time window. This incomplete decoupling again demonstrates that the low-temperature magnetic state is neither fully homogeneous nor purely static, but rather contains persistent dynamic magnetic regions.

The temperature dependence of the amplitudes A$_{1}$ and A$_{2}$ (shown in Fig. 7b) shows a strong redistribution around 5-8 K, where the slow-relaxing component A$_{2}$ increases rapidly while the fast-relaxing component A$_{1}$ decreases. This crossover suggests a gradual reorganization of the magnetic state rather than a sharp transition into a single uniform phase. The relaxation rates $\lambda$$_{1}$ and $\lambda$$_{2}$ shown in Fig. 7c, display clear maxima around 5-8 K, characteristic of the slowing down of spin fluctuations near the magnetic ordering region. Above about 15 K, the spectra become dominated by the slow-relaxing component, indicating that the internal fields become increasingly dynamic on the $\mu$SR timescale and indicating the spin reorientation, which has also been shown in the ZF data in Fig. 3a,b, although finite relaxation remains throughout the measured temperature range \cite{Ref18}. With further increase in temperature, both relaxation rates decrease gradually but remain finite even up to 60 K. The persistence of relaxation under a relatively strong LF indicates that correlated spin fluctuations survive over the entire temperature range.

There might be a concern regarding the apparent correlation between the fitted initial asymmetry parameters in Fig. 3(a) and Fig. 7(b), where A$_1$ and A$_2$ exhibit opposite trends over part of the temperature range. Such behavior is common in multi-component $\mu$SR analyses and has been reported in systems with multiple magnetic environments or muon stopping sites, where spectral weight is redistributed between relaxation components while the total observable asymmetry remains nearly conserved. Similar effects have been discussed in standard $\mu$SR literature and studies of magnetically ordered or phase-separated materials \cite{Ref37, Ref38}. In the present case, the correlation becomes noticeable above $\sim$ 18 K, within the temperature range where a gradual spin reorientation is expected. As the system evolves between the magnetic phases associated with the transitions at 46 K and 10 K, the local field distribution at the muon sites changes continuously, leading to a redistribution of muons among distinct magnetic environments and the observed complementary variation of A$_1$ and A$_2$. Therefore, these amplitude changes are more naturally interpreted as a redistribution of the relative fractions of muons sensing different local magnetic environments, consistent with the evolution of magnetic correlations and the proposed spin-reorientation process, rather than as a loss of muon polarization. Importantly, our conclusions are based on the temperature dependence of the relaxation rates and local field distributions, which show clear anomalies at the magnetic transitions and remain robust within fitting uncertainties. Thus, the partial covariance between A$_1$ and A$_2$ does not affect the interpretation of the magnetic behavior discussed in this work.

Collectively, the LF $\mu$SR measurements at 2000 G reveal the coexistence of quasi-static and dynamic magnetic components together with a broad distribution of fluctuation timescales. The incomplete field decoupling, spectral weight transfer between the two relaxation components, and persistent finite relaxation strongly support the existence of an inhomogeneous magnetic ground state in Ba$_{3}$HoRu$_{2}$O$_{9}$.

\textcolor{black}{As established in our ZF- and LF-$\mu$SR analyses, the asymmetry spectra for both Ba$_3$HoRu$_2$O$_9$ and Ba$_3$GdRu$_2$O$_9$ are well described by multi-component relaxation functions. In complex 6H-perovskite structures, the presence of multiple relaxation components naturally suggests that the implanted muons occupies two or more inequivalent crystallographic stopping sites. To rigorously explore this geometric possibility and distinguish it from magnetic ground state's inhomogeneity, we performed muon stopping site and electronic dipolar field calculations using machine-learning-based software, MEOWN \cite{ref35}.} 

\begin{figure*}[htbp]
  \centering
  \includegraphics[width=\textwidth]{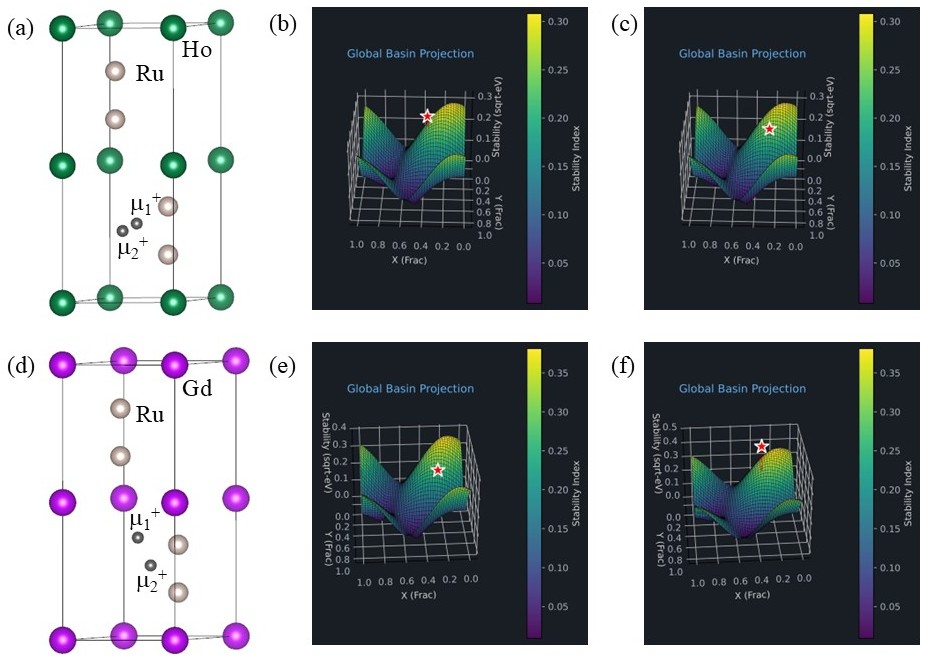}
  \caption{\textcolor{black}{Machine-learning-assisted muon stopping site predictions and electrostatic potential landscapes. (a) The crystal structure of Ba$_{3}$HoRu$_{2}$O$_{9}$ illustrating the two predicted muon stopping sites, denoted as $\mu_1^+$ and $\mu_2^+$. (b, c) The corresponding electrostatic energy global basin projections for (b) site 1 and (c) site 2 in Ba$_{3}$HoRu$_{2}$O$_{9}$, demonstrating their structural stability. (d) The crystal structure of Ba$_{3}$GdRu$_{2}$O$_{9}$ displaying its two calculated muon stopping sites, $\mu_1^+$ and $\mu_2^+$. (e, f) The corresponding electrostatic energy global basin projections for (e) site 1 and (f) site 2 in Ba$_{3}$GdRu$_{2}$O$_{9}$. The red stars in the basin projections represent the localized potential minima corresponding to the stabilized muon positions.}}
  \label{Fig-8}
\end{figure*}

\textcolor{black}{The calculations were performed on an $80 \times 80 \times 80$ voxel grid resolution to accurately map the electrostatic potential landscape of the unit cell. The optimization process utilized 5000 training iterations, followed by 500 L-BFGS-B steps for local minimization, and 500 Newton-Raphson steps for structural relaxation.} 

\textcolor{black}{For Ba$_3$HoRu$_2$O$_9$, previous neutron diffraction study reveals the microscopic coexistence of two distinct magnetic structures below $T_{N2}$ associated with the propagation vectors $K_1 = (0.5\ 0\ 0)$ and $K_2 = (0.25\ 0.25\ 0)$ \cite{Ref18}. To quantitatively evaluate the local magnetic environments, we calculated the dipolar fields for both scenarios using their respective saturation effective magnetic moments. For the $K_1$ phase (Ho = $4.952\ \mu_B$, Ru = $0.524\ \mu_B$), the algorithm identifies two stable muon stopping sites: the first at fractional coordinates (0.269, 0.356, 0.275) corresponding to the Wyckoff $24l$ position with a calculated dipolar field of 1678.3~G, and the second at (0.368, 0.272, 0.250) on the Wyckoff $12j$ position with a field of 1447.8~G. When evaluating the $K_2$ phase (Ho = $4.127\ \mu_B$, Ru = $1.105\ \mu_B$), the spatial muon coordinates remain structurally identical, but the differing magnetic spin texture yields significantly altered dipolar fields of 1273~G for the $24l$ site and 776.8~G for the $12j$ site. These predicted muon sites embedded within the crystal lattice, along with the electrostatic field stability plots, are illustrated in Fig. 9.}

\textcolor{black}{For the isostructural Ba$_3$GdRu$_2$O$_9$ compound, utilizing the saturation moments of Gd = $7.94\ \mu_B$ and Ru = $1.7\ \mu_B$, the model also predicts two distinct stopping sites. Both are located at general $24l$ Wyckoff positions, with coordinates at (0.250, 0.375, 0.260) and (0.322, 0.300, 0.362), yielding much larger static dipolar fields of 3247.9~G and 5095.2~G, respectively. This larger and more discrete field scale aligns perfectly with the homogenous, quasi-static Gaussian relaxation observed experimentally in the Gd system.}

\textcolor{black}{The identification of multiple stopping sites with structural multiplicities $24l$ and \ $12j$ in the Ho compound naturally explains the baseline origin of the two-component exponential relaxation. However, the purely geometric multi-site scenario does not account for the persistent spin dynamics. The calculated dipolar fields across the coexisting $K_1$ and $K_2$ domains span a broad range (from $\sim 776$~G to $\sim 1678$~G), providing excellent quantitative agreement with the dynamic root-mean-square local field ($H_{\text{loc}} \sim 1412$--$1488$~G) extracted from our Redfield LF decoupling analysis. Therefore, while the crystal geometry dictates the presence of the two relaxation channels, the heavily distributed internal fields and the incomplete LF decoupling up to 46~K are direct microscopic consequences of muons sampling these coexisting $K_1$ and $K_2$ magnetic phases. This provides robust physical justification that the multi-scale complex magnetism in Ba$_3$HoRu$_2$O$_9$ represents a genuinely inhomogeneous magnetic ground state, rather than a mere structural artifact.}

\textcolor{black}{Taken together, the ZF, field-dependent LF, and fixed-field $\mu$SR measurements establish that Ba$_3$HoRu$_2$O$_9$ hosts a multi-scale magnetic state where quasi-static magnetism and persistent spin dynamics coexist over a broad temperature range. In contrast, Ba$_3$GdRu$_2$O$_9$ settles into a much simpler, homogeneous magnetic state, entirely consistent with the isotropic, $L=0$ character of the Gd$^{3+}$ moments.} 

\textcolor{black}{The different magnetic behavior between heavy rare-earth elements (Ho, Dy) and Gd might arise due to the contrasting isotropic nature of Gd$^{3+}$ (L=0) compared to the high anisotropy of other rare-earth members. For example, in the R$_2$BaNiO$_5$ family, the isotropic Gd compound behaves differently from the anisotropic ones (like Dy, Er, Nd) \cite{ref39}.  However, this simple view falls short for many strongly correlated $d$-$f$ systems which show versatile magnetic behavior. For well-known multiferroic RMn$_2$O$_5$, the various R-members exhibit intriguing magnetic ground states, and the isotropic nature of Gd plays only a nominal role. For GdMn$_2$O$_5$, strong exchange frustration due to various competing interactions plays a crucial role in deciding the magnetic ground state, which shows large ferroelectricity \cite{ref40}. Specifically, in the Ba$_3$LnRu$_2$O$_9$ family, it is already documented that the 4$d$-4$f$ magnetic coupling and competing interactions plays the dominant role. For example, in our Ba$_3$RRu$_2$O$_9$ series, the magnetocaloric responses of Ho, Tb, and the isotropic Gd are actually very similar, while the Nd compound acts entirely differently \cite{Ref22}. Furthermore, the compound Ba$_3$HoRu$_2$O$_9$ is unique in its family, showing spin-driven ferroelectricity and two competing magnetic phases at low temperature, which is not observed for other members \cite{Ref20}. Therefore, the inhomogeneous magnetic state we observe is not just a basic result of Ho being anisotropic. Instead, it is driven by competing 4$d$-4$f$ interactions that frustrate the magnetic spins at a local level as well. To capture these complex, multi-scale spin dynamics, we need the deep microscopic sensitivity of $\mu$SR, a technique rarely applied to this extent in $d$-$f$ coupled oxides.}

\textcolor{black}{To physically ground this observation, we employed the MEOWN software framework to map the muon stopping sites and calculate the local electronic dipolar fields. The integration of machine-learning-assisted algorithms for muon site prediction and internal field modelling represents a novel and cutting-edge approach in modern $\mu$SR spectroscopy. These computational results perfectly complement our experimental conclusions, the wide distribution of internal magnetic fields measured by $\mu$SR is not a random anomaly, but naturally emerges from muons stabilized at inequivalent structural sites and sensing the distinct, coexisting $K_1$ and $K_2$ magnetic domains. The fact that these domains survive down to the lowest temperatures without fully freezing suggests a magnetic landscape dominated by intense local frustration.}

\textcolor{black}{To place these findings in a broader fundamental context, a landmark study by Zorko \textit{et al.} on $\alpha$-NaMnO$_2$ demonstrated that geometrical frustration can spontaneously generate nanoscale magnetic inhomogeneity in a chemically pristine, single-phase 3$d$ spin lattice \cite{ref41}. Our study conceptually enriches this paradigm by extending it into the highly complex territory of strongly correlated $d$-$f$ systems. Such a unique phenomenon is not reported in any d-f coupled magnetic systems, including well-studied RMnO$_3$, RMn$_2$O$_5$, RCrTiO$_5$ \cite{ref42, ref43, ref44, ref45, ref46}. In Ba$_3$HoRu$_2$O$_9$, the inhomogeneity does not stem from simple geometric frustration on a triangular lattice, but rather from the fierce energetic competition between distinct 4$d$ (Ru) and 4$f$ (Ho) magnetic exchange pathways. This competition is heavily amplified by the strong crystal-field-induced rare earth anisotropy of Ho$^{3+}$. We propose that this delicate energetic balance prevents the spin lattice from settling into a single, uniform ordered state. Instead, much like the 3$d$ frustrated systems but driven by a completely different 4$d$--4$f$ mechanism, the material relieves its magnetic degeneracy by going into a dynamic, phase-coexisting landscape. Changing the complex Ho ion to the simpler isotropic Gd ion makes the magnetic state uniform again.}

\section{Conclusion}

The $\mu$SR investigation on Ba$_{3}$HoRu$_{2}$O$_{9}$ and Ba$_{3}$GdRu$_{2}$O$_{9}$ provides a microscopic understanding of how rare-earth anisotropy controls the magnetic ground state and spin dynamics in these 4$d$--4$f$ coupled systems. Although both compounds belong to the same Ba$_{3}$LnRu$_{2}$O$_{9}$ family and share similar crystal structures, they exhibit fundamentally different magnetic behavior, demonstrating that the magnetic ground state can be tuned by changing the rare-earth ion.

The ZF, LF, and fixed LF-field temperature dependent $\mu$SR measurements establish that Ba$_{3}$HoRu$_{2}$O$_{9}$ does not develop a simple homogeneous magnetic ground state. Instead, it exhibits coexistence of quasi-static magnetism and persistent spin dynamics over a broad temperature range. The two-component exponential relaxation, absence of coherent oscillation below the ordering temperature, and incomplete LF decoupling indicate a broad distribution of local magnetic environments and fluctuation timescales. The $\mu$SR results further reveal two magnetic energy scales, with correlated magnetic behavior persisting up to nearly 46 K and quasi-static ordering developing around 10 K. The Redfield analysis shows that the local magnetic field remains large even above the ordering temperature, while the correlation time decreases systematically with increasing temperature. This suggests that short-range magnetic correlations and dynamic spin fluctuations coexist over a wide temperature range, which may result in an inhomogeneous and multi-scale magnetic ground state in Ba$_{3}$HoRu$_{2}$O$_{9}$. \textcolor{black} {To physically support these experimental findings, our machine-learning-assisted MEOWN software's calculations successfully mapped the muon stopping sites and local dipolar field. This modern computational approach shows that the broad field distribution naturally comes from muons sensing the coexisting magnetic domains at different structural sites, strongly backing our conclusion of a highly frustrated and inhomogeneous state.} In contrast, Ba$_{3}$GdRu$_{2}$O$_{9}$ exhibits a comparatively homogeneous antiferromagnetic state below 14.5 K. The coexistence of Gaussian and exponential relaxation components indicates a crossover from a dynamic paramagnetic state to a relatively uniform static magnetic ground state, consistent with the nearly isotropic nature of the Gd$^{3+}$ ($L = 0$) moments.

The comparison between the Ho and Gd-based compounds clearly shows that the magnetic ground state in the Ba$_{3}$LnRu$_{2}$O$_{9}$ series can be tuned from homogeneous to inhomogeneous by changing the rare-earth ion. The strong spin-orbit coupling and rare earth anisotropy of Ho$^{3+}$ stabilize distributed magnetic dynamics, whereas the isotropic Gd$^{3+}$ moments favor conventional magnetic order. \textcolor{black} {Changing the complex Ho ion to the simpler Gd ion makes the magnetic state uniform again. This shows that the unique interaction between the 4$d$ and 4$f$ electrons is responsible for the complex behavior. In the end, this study gives a clear, atomic-level view of how we can use different rare-earth ions and competing magnetic forces as ``control knobs'' to tune and manage complex magnetic states in advanced materials, which is highly relevant for future spintronic and quantum applications.}

\section{acknowledgments}
T.B. greatly acknowledges the Science and Engineering Research Board (SERB), now, Anusandhan National Research Foundation (ANRF) (Project No.: SRG/2022/000044), Government of India, for research funding. SG thanks SERB (now, ANRF) for the fellowship. The authors thank the ISIS Facility, STFC, UK, for muon beam time on the EMU (Proposal no. RB2510553) \cite{ref47} and HIFI (Proposal no. XB2590019) \cite{ref48} instrument and the Department of Science and Technology, India, for financial support during the experiment. D.T.A. acknowledges EPSRC UK (Grant Ref: EP/W00562X/1) for funding.

\section*{Competing interests}
The authors declare no competing financial interests.

\bibliography{your_bib_file}

@article{Ref1,
  title={Designing and controlling the properties of transition metal oxide quantum materials},
  author={Ahn, Charles and Cavalleri, Andrea and Georges, Antoine and Ismail-Beigi, Sohrab and Millis, Andrew J and Triscone, Jean-Marc},
  journal={Nature materials},
  volume={20},
  number={11},
  pages={1462--1468},
  year={2021},
  publisher={Nature Publishing Group UK London}
}

@article{Ref2,
  title={Electric-field control of the orbital occupancy and magnetic moment of a transition-metal oxide},
  author={Preziosi, Daniele and Alexe, Marin and Hesse, Dietrich and Salluzzo, Marco},
  journal={Physical review letters},
  volume={115},
  number={15},
  pages={157401},
  year={2015},
  publisher={APS}
}

@article{Ref3,
  title={Transition metal oxides: a new frontier in spintronics driven by novel quantum states and efficient charge-spin interconversion},
  author={Han, Yamin and Lao, Bin and Zheng, Xuan and Li, Sheng and Li, Run-Wei and Wang, Zhiming},
  journal={Frontiers in Materials},
  volume={11},
  pages={1444769},
  year={2024},
  publisher={Frontiers Media SA}
}

@article{Ref4,
  title={Oxide spin-orbitronics: spin--charge interconversion and topological spin textures},
  author={Trier, Felix and No{\"e}l, Paul and Kim, Joo-Von and Attan{\'e}, Jean-Philippe and Vila, Laurent and Bibes, Manuel},
  journal={Nature Reviews Materials},
  volume={7},
  number={4},
  pages={258--274},
  year={2022},
  publisher={Nature Publishing Group UK London}
}

@article{Ref5,
  title={Unconventional S-orbital state of Tb and cooperative Ru (4d)--Tb (4f) spin-ordering in the strongly correlated 4d--4f system Ba 3 TbRu 2 O 9},
  author={Kushwaha, E and Roy, G and dos Santos, AM and Kumar, M and Ghosh, S and Heitmann, T and Basu, T},
  journal={Journal of Materials Chemistry C},
  volume={13},
  number={30},
  pages={15384--15389},
  year={2025},
  publisher={Royal Society of Chemistry}
}

@article{Ref6,
  title={Unconventional spin-intertwined charge density wave in magnetic phases of kagome metal GdTi3Bi4},
  author={Han, Xianghe and Chen, Hui and Cao, Zhongyi and Guo, Jingwen and Tuo, Chuyi and Li, Ming-Rui and Fei, Fucong and Tan, Hengxin and Guo, Jianfeng and Shi, Yanhao and others},
  journal={Nature Communications},
  year={2026},
  publisher={Nature Publishing Group UK London}
}

@article{Ref7,
  title={Unconventional magnetism in the 4 d 4-based S= 1 honeycomb system Ag 3 LiRu 2 O 6},
  author={Kumar, R and Dey, Tusharkanti and Ette, PM and Ramesha, K and Chakraborty, Atasi and Dasgupta, I and Orain, JC and Baines, C and T{\'o}th, S{\'a}ndor and Shahee, A and others},
  journal={Physical Review B},
  volume={99},
  number={5},
  pages={054417},
  year={2019},
  publisher={APS}
}

@article{Ref8,
  title={Stability of 4f configurations in rare-earth-metal compounds},
  author={Laubschat, C and Kaindl, G and Schneider, W-D and Reihl, B and Mrtensson, N},
  journal={Physical Review B},
  volume={33},
  number={10},
  pages={6675},
  year={1986},
  publisher={APS}
}

@article{Ref9,
  title={Magnetic circular dichroism of x-ray emission for gadolinium in 4d-4f excitation region},
  author={Takayama, Yasuhiro and Shinoda, Motoki and Obu, Kenji and Lee, Chol and Shiozawa, Hidetsugu and Hirose, Masaaki and Ishii, Hiroyoshi and Miyahara, Tsuneaki and Okamoto, Jun},
  journal={Journal of the Physical Society of Japan},
  volume={71},
  number={1},
  pages={340--346},
  year={2002},
  publisher={The Physical Society of Japan}
}

@article{Ref10,
  title={The coordination chemistry and magnetism of some 3d--4f and 4f amino-polyalcohol compounds},
  author={Sharples, Joseph W and Collison, David},
  journal={Coordination chemistry reviews},
  volume={260},
  pages={1--20},
  year={2014},
  publisher={Elsevier}
}

@article{Ref11,
  title={Tuning quantum tunnelling of magnetization through 3d--4f magnetic interactions: an alternative approach for manipulating single-molecule magnetism},
  author={Li, Jing and Wei, Rong-Min and Pu, Tian-Cheng and Cao, Fan and Yang, Li and Han, Yuan and Zhang, Yi-Quan and Zuo, Jing-Lin and Song, You},
  journal={Inorganic Chemistry Frontiers},
  volume={4},
  number={1},
  pages={114--122},
  year={2017},
  publisher={Royal Society of Chemistry}
}

@article{Ref12,
  title={Utilizing 3d--4f magnetic interaction to slow the magnetic relaxation of heterometallic complexes},
  author={Li, Xiao-Lei and Min, Fan-Yong and Wang, Chao and Lin, Shuang-Yan and Liu, Zhiliang and Tang, Jinkui},
  journal={Inorganic Chemistry},
  volume={54},
  number={9},
  pages={4337--4344},
  year={2015},
  publisher={ACS Publications}
}

@article{Ref13,
  title={Interplay between 3d--3d and 3d--4f interactions at the origin of the magnetic ordering in the Ba2LnFeO5 oxides},
  author={Kundu, Asish K and Hardy, Vincent and Caignaert, Vincent and Raveau, Bernard},
  journal={Journal of Physics: Condensed Matter},
  volume={27},
  number={48},
  pages={486001},
  year={2015},
  publisher={IOP Publishing}
}

@article{Ref14,
  title={Modification of magnetic anisotropy through 3d-4f coupling in La0. 75Pr0. 25Co2P2},
  author={Kovnir, Kirill and Thompson, Corey M and Garlea, V Ovidiu and Haskel, Daniel and Polyanskii, Anatolii A and Zhou, Haidong and Shatruk, Michael},
  journal={Phys. Rev. B},
  volume={88},
  number={10},
  pages={104429},
  year={2013}
}

@article{Ref15,
  title={Crystal structures and magnetic properties of the 6H-perovskites Ba3LnRu2O9 (Ln= Ce, Pr and Tb)},
  author={Doi, Yoshihiro and Wakeshima, Makoto and Hinatsu, Yukio and Tobo, Aya and Ohoyama, Kenji and Yamaguchi, Yasuo},
  journal={Journal of Materials Chemistry},
  volume={11},
  number={12},
  pages={3135--3140},
  year={2001},
  publisher={Royal Society of Chemistry}
}

@article{Ref16,
  title={Enhancement of magnetodielectric coupling in 6H-perovskites Ba 3 R Ru 2 O 9 for heavier rare-earth cations (R= Ho, Tb)},
  author={Basu, Tathamay and Caignaert, Vincent and Ghara, Somnath and Ke, Xianglin and Pautrat, Alain and Krohns, Stephan and Loidl, Alois and Raveau, Bernard},
  journal={Physical Review Materials},
  volume={3},
  number={11},
  pages={114401},
  year={2019},
  publisher={APS}
}

@article{Ref17,
  title={Structures and magnetic properties of double perovskites A2LnMO6 and 6H-perovskites Ba3LnRu2O9 (A= Sr, Ba; Ln= Y, lanthanides; M= Nb, Ta, Ru)},
  author={Hinatsu, Yukio and Doi, Yoshihiro},
  journal={Bulletin of the Chemical Society of Japan},
  volume={76},
  number={6},
  pages={1093--1113},
  year={2003},
  publisher={Oxford University Press}
}

@article{Ref18,
  title={Cooperative Ru (4 d)-Ho (4 f) magnetic ordering and phase coexistence in the 6 H perovskite multiferroic Ba 3 Ho Ru 2 O 9},
  author={Basu, T and Caignaert, Vincent and Damay, F and Heitmann, TW and Raveau, B and Ke, X},
  journal={Physical Review B},
  volume={102},
  number={2},
  pages={020409},
  year={2020},
  publisher={APS}
}

@article{Ref19,
  title={The structural and magnetic characterization of 6H-perovskite-type oxides Ba3LnIr2O9 (Ln= Y, lanthanides)},
  author={Doi, Yoshihiro and Hinatsu, Yukio},
  journal={Journal of Physics: Condensed Matter},
  volume={16},
  number={16},
  pages={2849--2860},
  year={2004}
}

@article{Ref20,
  title={Origin of spin-driven ferroelectricity and effect of external pressure on the complex magnetism of the 6 H perovskite Ba 3 Ho Ru 2 O 9},
  author={Kushwaha, Ekta and Roy, Gourab and Kumar, Mohit and dos Santos, Antonio Moreira and Ghosh, Sayan and Adroja, DT and Caignaert, Vincent and Perez, O and Pautrat, Alain and Basu, Tathamay},
  journal={Physical Review B},
  volume={109},
  number={22},
  pages={224418},
  year={2024},
  publisher={APS}
}

@article{Ref21,
  title={Magnetic and calorimetric studies on Ba 3 LnRu 2 O 9 (Ln= Gd, Ho--Yb) with 6H-perovskite structure},
  author={Doi, Yoshihiro and Hinatsu, Yukio},
  journal={Journal of Materials Chemistry},
  volume={12},
  number={6},
  pages={1792--1795},
  year={2002},
  publisher={Royal Society of Chemistry}
}

@article{Ref22,
  title={Intriguing Magnetocaloric Effect in Multiferroic Ba3RRu2O9 (R= Ho, Gd, Tb, Nd) with Strong 4d--4f Correlations},
  author={Kumar, Mohit and Ghosh, Sayan and Roy, Gourab and Kushwaha, Ekta and Caignaert, Vincent and Prellier, Wilfrid and Majumdar, Subham and Hardy, Vincent and Basu, Tathamay},
  journal={ACS Applied Energy Materials},
  year={2026},
  publisher={ACS Publications}
}

@article{Ref23,
  title={Exploring a New Regime of Molecular Orbital Physics in 4 d Cluster Magnets with Resonant Inelastic X-Ray Scattering},
  author={Yuan, Bo and Kim, Beom Hyun and Chen, Qiang and Dobrowolski, Daniel and Azmanska, Monika and Luke, GM and Fan, Shiyu and Bisogni, Valentina and Pelliciari, Jonathan and Clancy, JP},
  journal={Physical Review Letters},
  volume={134},
  number={10},
  pages={106504},
  year={2025},
  publisher={APS}
}

@article{Ref24,
  title={Frustrated spin-1 2 molecular magnetism in the mixed-valence antiferromagnets Ba 3 M Ru 2 O 9 (M= In, Y, Lu)},
  author={Ziat, D and Aczel, Adam A and Sinclair, R and Chen, Q and Zhou, HD and Williams, Travis J and Stone, Matthew B and Verrier, A and Quilliam, JA},
  journal={Physical Review B},
  volume={95},
  number={18},
  pages={184424},
  year={2017},
  publisher={APS}
}

@article{Ref25,
  title={Spin orders and lattice distortions of geometrically frustrated 6H-perovskites Ba 3 B' Ru 2 O 9 (B'= La 3+, Nd 3+, and Y 3+)},
  author={Senn, Mark S and Kimber, Simon AJ and Arevalo Lopez, Angel M and Hill, Adrian H and Attfield, J Paul},
  journal={Physical Review B—Condensed Matter and Materials Physics},
  volume={87},
  number={13},
  pages={134402},
  year={2013},
  publisher={APS}
}

@article{Ref26,
  title={Realization of the orbital-selective Mott state at the molecular level in Ba 3 LaRu 2 O 9},
  author={Chen, Q and Verrier, A and Ziat, D and Clune, AJ and Rouane, R and Bazier-Matte, X and Wang, G and Calder, Stuart and Taddei, Keith M and Cruz, CR dela and others},
  journal={Physical Review Materials},
  volume={4},
  number={6},
  pages={064409},
  year={2020},
  publisher={APS}
}

@article{Ref27,
  title={Magnetodielectric coupling in a Ru-based 6H-perovskite, Ba3NdRu2O9},
  author={Basu, Tathamay and Pautrat, Alain and Hardy, Vincent and Loidl, Alois and Krohns, Stephan},
  journal={Applied Physics Letters},
  volume={113},
  number={4},
  year={2018},
  publisher={AIP Publishing}
}

@article{Ref28,
  title={Muon spin spectroscopy},
  author={Hillier, Adrian D and Blundell, Stephen J and McKenzie, Iain and Umegaki, Izumi and Shu, Lei and Wright, Joseph A and Prokscha, Thomas and Bert, Fabrice and Shimomura, Koichiro and Berlie, Adam and others},
  journal={Nature Reviews Methods Primers},
  volume={2},
  number={1},
  pages={4},
  year={2022},
  publisher={Nature Publishing Group UK London}
}

@book{Ref29,
  title={Muon spectroscopy: an introduction},
  author={Blundell, Stephen and De Renzi, Roberto and Lancaster, Tom and Pratt, Francis L},
  year={2021},
  publisher={Oxford University Press}
}

@book{Ref30,
  title={Muon spin rotation, relaxation, and resonance: applications to condensed matter},
  author={Le Yaouanc, Alain and De Reotier, Pierre Dalmas},
  number={147},
  year={2011},
  publisher={OUP Oxford}
}

@article{Ref31,
  title={Optimising a muon spectrometer for measurements at the ISIS pulsed muon source},
  author={Giblin, SR and Cottrell, SP and King, PJC and Tomlinson, S and Jago, SJS and Randall, LJ and Roberts, MJ and Norris, J and Howarth, S and Mutamba, QB and others},
  journal={Nuclear Instruments and Methods in Physics Research Section A: Accelerators, Spectrometers, Detectors and Associated Equipment},
  volume={751},
  pages={70--78},
  year={2014},
  publisher={Elsevier}
}

@article{Ref32,
  title={Design and commissioning of a high magnetic field muon spin relaxation spectrometer at the ISIS pulsed neutron and muon source},
  author={Lord, JS and McKenzie, I and Baker, PJ and Blundell, SJ and Cottrell, SP and Giblin, SR and Good, J and Hillier, AD and Holsman, BH and King, PJC and others},
  journal={Review of Scientific Instruments},
  volume={82},
  number={7},
  year={2011},
  publisher={AIP Publishing}
}

@article{Ref33,
  title={Interplay between trimer structure and magnetic ground state in Ba 5 Ru 3 O 12 probed by neutron and $\mu$ SR techniques},
  author={Kushwaha, E and Ghosh, S and Sannigrahi, J and Roy, G and Kumar, M and Cottrell, S and Stone, MB and Fang, Y and Adroja, DT and Ke, X and others},
  journal={Physical Review B},
  volume={112},
  number={9},
  pages={094410},
  year={2025},
  publisher={APS}
}

@article{Ref34,
  title={WIMDA: a muon data analysis program for the Windows PC},
  author={Pratt, FL},
  journal={Physica B: Condensed Matter},
  volume={289},
  pages={710--714},
  year={2000},
  publisher={Elsevier}
}

@article{ref35,
  title={Development of a Physics-Informed Neural Framework, {MEOWN}, for Rapid Prediction of Muon Stopping Sites in Crystalline Materials, for understanding Quantum Magnet employing Muon Spectroscopy},
  author={Pandey, A. and Sharma, K. and Ghosh, S. and Roy, G. and Basu, T.},
  journal={arXiv preprint arXiv:2609.17063},
  year={2026},
  note={The code and software are protected by a copyright agreement (SW-2026023675, Copyright Office, Govt. of India)}
}

@book{Ref36,
  author    = {Charles P. Slichter},
  title     = {Principles of Magnetic Resonance},
  publisher = {Springer},
  edition   = {3},
  year      = {1990},
  series    = {Springer Series in Solid-State Sciences},
  volume    = {1},
  address   = {Berlin, Heidelberg},
}

@article{Ref37,
  title={Muon spin relaxation and neutron diffraction investigations of quadrupolar and magnetically ordered states of YbRu 2 Ge 2},
  author={Jeevan, HS and Adroja, DT and Hillier, AD and Hossain, Z and Ritter, C and Geibel, C},
  journal={Physical Review B—Condensed Matter and Materials Physics},
  volume={84},
  number={18},
  pages={184405},
  year={2011},
  publisher={APS}
}

@article{Ref38,
  title={Muon spin rotation study of the layered oxyselenide Sr 2 CoO 2 Ag 2 Se 2},
  author={Gill, GJW and Wilkinson, JM and Lang, F and Blandy, JN and Clarke, SJ and Blundell, SJ},
  journal={Physical Review B},
  volume={106},
  number={22},
  pages={224410},
  year={2022},
  publisher={APS}
}

@article{ref39,
  title={Spin-Chain Incipient Magnetocaloric Effect and Rare-Earth ({R}) Controlled Switching in the Haldane-Chain System, {R$_2$BaNiO$_5$}},
  author={Kumar, M. and Roy, G. and Ghosh, S. and Kushwaha, E. and Singh, K. and Basu, T.},
  journal={Advanced Quantum Technologies},
  volume={9},
  number={6},
  pages={e70356},
  year={2026},
  publisher={Wiley Online Library}
}

@article{ref40,
  title={Giant tunability of ferroelectric polarization in {GdMn$_2$O$_5$}},
  author={Lee, N. and Vecchini, C. and Choi, Y. J. and Chapon, L. C. and Bombardi, A. and Radaelli, P. G. and Cheong, S. W.},
  journal={Physical Review Letters},
  volume={110},
  number={13},
  pages={137203},
  year={2013},
  publisher={APS}
}

@article{ref41,
  title={Frustration-induced nanometre-scale inhomogeneity in a triangular antiferromagnet},
  author={Zorko, A. and Adamopoulos, O. and Komelj, M. and Ar{\v{c}}on, D. and Lappas, A.},
  journal={Nature Communications},
  volume={5},
  number={1},
  pages={3222},
  year={2014},
  publisher={Nature Publishing Group UK London}
}

@article{ref42,
  title={Local magnetism and magnetoelectric effect in {HoMnO$_3$} studied with muon-spin relaxation},
  author={Lewtas, H. J. and Lancaster, T. and Baker, P. J. and Blundell, S. J. and Prabhakaran, D. and Pratt, F. L.},
  journal={Physical Review B},
  volume={81},
  number={1},
  pages={014402},
  year={2010},
  publisher={APS}
}

@article{ref43,
  title={Magnetism in geometrically frustrated {YMnO$_3$} under hydrostatic pressure studied with muon spin relaxation},
  author={Lancaster, T. and Blundell, S. J. and Andreica, D. and Janoschek, M. and Roessli, B. and Gvasaliya, S. N. and others},
  journal={Physical Review Letters},
  volume={98},
  number={19},
  pages={197203},
  year={2007},
  publisher={APS}
}

@article{ref44,
  title={Investigation of the {GdMn$_2$O$_5$} multiferroic by the {$\mu$SR} method},
  author={Vorob'ev, S. I. and Getalov, A. L. and Golovenchits, E. I. and Komarov, E. N. and Koptev, V. P. and Kotov, S. A. and others},
  journal={Physics of the Solid State},
  volume={55},
  number={3},
  pages={466--475},
  year={2013},
  publisher={Springer}
}

@article{ref45,
  title={Muon-spin-relaxation investigation of {EuMn$_2$O$_5$}},
  author={Vorob'ev, S. I. and Golovenchits, E. I. and Koptev, V. P. and Komarov, E. N. and Kotov, S. A. and Sanina, V. A. and Shcherbakov, G. V. E.},
  journal={JETP letters},
  volume={91},
  number={10},
  pages={512--517},
  year={2010},
  publisher={Springer}
}

@article{ref46,
  title={Complex nature of magnetic field induced ferroelectricity in {GdCrTiO$_5$}},
  author={Basu, T. and Adroja, D. T. and Kolb, F. and Krug von Nidda, H. A. and Ruff, A. and Hemmida, M. and others},
  journal={Physical Review B},
  volume={96},
  number={18},
  pages={184431},
  year={2017},
  publisher={APS}
}

@misc{ref47, 
title={Muon Spin Relaxation Study of the Complex Magnetic Ground States of a Ruthenate Dimer System}, url={https://data.isis.stfc.ac.uk/doi/INVESTIGATION/128218602/}, 
DOI={10.5286/ISIS.E.RB2510553-1}, publisher={ISIS Facility}, 
author={D. T. Adroja}, 
year={2025} 
}

@misc{ref48, 
title={Investigation of Ground State of Ru and Gd+3 ions (S-state lanthanide ion) in Ba3GdRu2O9}, url={}, 
DOI={}, publisher={ISIS Facility}, 
author={Tathamay Basu}, 
year={2025} 
}

\end{document}